\documentclass[preprint,prd,nofootinbib]{revtex4-1}
\usepackage{color}
\usepackage{graphicx}
\usepackage{epsfig}
\usepackage{subfig}
\usepackage{ctable}
\usepackage{hyperref}
\usepackage{physics}
\usepackage{graphicx}
\usepackage{dcolumn}
\usepackage{bm}

\begin{document}

 \begin{center}
     \Large{\textbf{ Noether Symmetry Approach to  the Non-Minimally Coupled $Y(R)F^2$ Gravity  }} \\[0.3cm]

     \large{\"{O}zcan SERT$^{\rm a}$,  Fatma \c{C}EL\.{I}KTA\c{S}$^{\rm b}$}
        \\[0.3cm]

        \small{
            \textit{$^{\rm a}$ Department of Physics,  Pamukkale University,  20017 Denizli, Turkey}   \\
              \textit{$^{\rm b}$ Department of Mathematics, Pamukkale University, 20017 Denizli, Turkey} 
            }

    \end{center}


\vspace{.6cm}

\hrule \vspace{0.2cm}
\noindent \small{\textbf{Abstract}

We use  Noether symmetry approach  to find spherically symmetric static solutions of  the non-minimally coupled electromagnetic fields to gravity.
We construct  the point-like  Lagrangian under the spherical symmetry assumption. Then we determine Noether symmetry and the corresponding conserved charge.   We derive Euler-Lagrange equations from this point-like Lagrangian
and show that these equations  are same with  the differential equations derived from the field equations of the model.
Also we give two  new exact asymptotically flat solutions  to these equations and investigate some  thermodynamic properties  of  these black holes.

\section{Introduction}
The late time expansion   of the universe and the missing matter  at large astrophysical scales  
still remain  the most important issues   of modern cosmology. Although Einstein's gravity with the cosmological constant known as $\Lambda$CDM  is in agreement with the  observations,  it  has some problems such as  fine tuning and coincidence. Since dark matter particle has not yet been observed directly, 
the efforts to  modify the Einstein's theory of gravity  have  recently increased by  studying  the models such as  $f(R)$ gravity, scalar-tensor gravity and vector-tensor gravity.

Due to the modification of the Einstein's gravity,
 it is possible  to  modify the  Einstein-Maxwell theory  to the  $f(R)$-Maxwell  theory in the presence of the electromagnetic field.
The spherically symmetric static solution  of this   theory  with constant Ricci scalar,  which  is similar to the Reissner-Nordstrom-AdS black hole solution, was given in \cite{Dombriz2009}.
However, in general cases with dynamical  non-constant Ricci scalar,  it is not easy to find  more general solutions.  Then one can take into account  the non-minimal couplings between electromagnetic fields and gravity like $Y(R)F^2$-type.  
The feature of the non-minimal theory is to have a large class of solutions 
such as  the spherically symmetric \cite{DereliSert20111,DereliSert20112,Sert2012Plus,Sert2013MPLA,Sert2016regular} and cosmological solutions \cite{AADS,Sert2019,SertAdak2019}. 
It is interesting to note that such non-minimal couplings, which are first order in R, have been obtained in \cite{MullerHoissen19882,MullerHoissen19883,Buchdahl1979,Dereli1990} from a five-dimensional Lagrangian via dimensional reduction. Also, \cite{Prasanna1971,Horndeski1976,MullerHoissen1988,Drummond1980} investigated various aspects of these couplings, such as charge conservation and the relationship between electric charge and geometry.
The general couplings in $R^nF^2$-form   applied to the generation of primordial magnetic fields in the inflation stage \cite{Turner1988,Campanelli2008,Kunze2010,Mazzi1995,AADS,Sert2019,Bamba2008}.
Therefore, it is possible  to consider the  more general couplings in $Y(R)F^2$ form and their solutions  can  give us more information about the relation between electric charge and space-time curvature. Especially in the presence of medium with very high density electromagnetic fields, these couplings may arise and  their effects can be significant even   far from the source.

The Noether symmetry approach is one of the effective techniques  to find  solutions of  a  Lagrangian without using  field equations.   This symmetry approach  allows us to find conserved quantities of a model by using  the  symmetry of the   Lagrangian which is invariant  along  a vector field.  Then the vector field can be determined by     this symmetry  and  each symmetry of the Lagrangian gives a conserved quantity. This symmetry approach has been applied to $f(R)$ gravity successfully
to find out solutions and select the corresponding  $f(R) $ function which is compatible with  the Noether symmetry \cite{Capozziello2007,Capozziello2000,Capozziello2008}.

This study is organized as follows:  In the second section,  we   find the first order point-like action of the model for the spherically symmetric static metric and  electric field.  After we apply the Noether symmetry approach to the action, we obtain the system of  partial differential equations. By solving  the system, we find  the Noether charge  and  the corresponding vector field for the non-minimal $Y(R)F^2$ model. In the  third section, after  we  obtain  Euler-Lagrange equations   we  give two new solutions to the equations and investigate some  thermodynamic properties of the black hole solutions. Finally, we  summarize  the results in the last section.

\section{Noether Symmetry Approach for the Non-Minimal Model} \label{model}
Let us  start with the following  action of the non-minimally coupled electromagnetic fields to gravity \cite{DereliSert20111,DereliSert20112,Bamba2008}
\begin{eqnarray}\label{eylemfon0}
\mathcal{I}=\int  \ \left[\frac{R}{2 \kappa ^2}*1-Y(R)F\wedge *F  \right] .
\end{eqnarray}
By taking the  variation of the action, and obtaining the field equations, we can find  the  solutions \cite{DereliSert20111,DereliSert20112,Sert2012Plus,Sert2013MPLA,Sert2016regular} for the spherically symmetric static metric 
 \begin{eqnarray}\label{metrik1}
 ds^2  = - A(r)dt^2 + \frac{1}{A(r)} dr^2 + B(r) d\theta^2+B(r) \sin^2\theta d\varphi^2 \ .  
 \end{eqnarray}
 Here the corresponding  Ricci curvature scalar is
 \begin{eqnarray}\label{R}
 R=-A''-\frac{2AB''}{B}- \frac{2A'B'}{B} +\frac{AB'^2}{ 2B^2 } +\frac{2}{B }\ .
 \end{eqnarray}

Alternatively,  we can find the solutions also from Noether symmetry approach by taking  the following  action of the  non-minimally coupled model  with the Lagrange multiplier $\lambda$ 
\begin{eqnarray}\label{eylemfon1}
\mathcal{I}=\int  \ \left[\frac{R}{2 \kappa ^2}*1-Y(R)F\wedge *F-\
\lambda (R-\bar{R})*1   \right] .
\end{eqnarray}
Here  variation of the action  with respect to $\lambda $ gives us $R=\bar{R}$ and $\bar{R}$  is  defined as
 \begin{eqnarray}\label{542}
 \bar{R}=R^* -A''-\frac{2AB''}{B} 
 \end{eqnarray}
to  eliminate  the second order derivatives in the action  via integration by parts. Here $R^*$ is defined as
$
R^* = - \frac{2A'B'}{B} +\frac{AB'^2}{ 2B^2 } +\frac{2}{B }\;.
$
The variation of the action  with respect to $R$ gives
 \begin{eqnarray} \label{543}
 \lambda =\frac{1}{2 \kappa ^2}*1 - Y_R (R)F\wedge *F   \ 
 \end{eqnarray}
 where $Y_R(R) = \dv{ Y(R)}{R}  $.
 If we substitute  (\ref{542}) and (\ref{543}) in the action (\ref{eylemfon1}), we obtain the following Lagrangian
 \begin{eqnarray}\label{544}
 L&=& 	 \frac{R}{2 \kappa ^2}*1- Y(R)F\wedge *F
 -\left(\frac{1}{2 \kappa ^2}*1- Y_R (R) F\wedge *F\right)\left(R-R^* +A''+\frac{2AB''}{B}\right) \ .
 \end{eqnarray}
We see that   $Y_R (R) F\wedge *F$ term in the Lagrangian  has    higher order derivatives  which complicates the Noether approach. But, fortunately we have  the following equation from the trace of the field equations
 \begin{eqnarray}\label{565}
 Y_R (R)F\wedge *F= - \frac{1}{2 \kappa^2}
 \end{eqnarray}
   which corresponds to the conservation  of the energy-momentum tensor \cite{Sert2019}  and 
  eliminates the higher order derivatives  in the Lagrangian.  By taking the electromagnetic tensor $F$, 
 \begin{eqnarray}\label{elektensor1}
 F = \phi'(r) e^1\ \wedge e^0 \;,
 \end{eqnarray}
  which has only the electric potential   $\phi (r)$, the  Lagrangian  of the model is obtained as 
  \begin{eqnarray}\label{541}
 L&=& 	 \frac{B}{ \kappa ^2} \left [-\frac{R}{2 }+\kappa ^2Y(R)\phi '^2+R^* -A''-\frac{2AB''}{B}\right] \;.
 \end{eqnarray}
 In the Lagrangian, the second order derivatives can be eliminated by   integration by parts and it turns out to be the following point-like Lagrangian
 \begin{eqnarray}\label{553}
 L =\frac{AB'^2}{2 \kappa^2 B} +\frac{A' B'}{\kappa^2}-\frac{BR}{2 \kappa^2}+BY(R)\phi '^2+\frac{2}{\kappa^2} \; . 
 \end{eqnarray} 
 By considering  the the  configuration space $Q$ which  has the generalized coordinates $q^i \equiv \{ A,  B , \phi  ,R \} $ 
 and       its tangent space $TQ \equiv \{ q^i, q'^i\} $,  we look for the symmetries of the Lagrangian.  Noether's theorem states that  if the Lie derivative of a Lagrangian   vanishes
\begin{eqnarray}
 \mathcal{L}_X L(q^i ,  q'^i)  = X L(q^i ,  q'^i)  =0
\end{eqnarray} along a vector field $X $ 
\begin{equation}
X = \alpha_i\frac{\partial}{\partial q^i} + \alpha'_i\frac{\partial}{\partial q'^i},
\end{equation}
then $X$ is  a symmetry  of the action and each symmetry of the action corresponds to a conserved quantity or first integral  such as
\begin{eqnarray}
\Sigma_0 =\alpha_i \frac{\partial L}{\partial q'^i} \;. \label{canstantofmotion}
\end{eqnarray}
Then we take the Lie derivative of the point-like  Lagrangian in the configuration space to find the first integral
\begin{eqnarray}
 \mathcal{L}_X L&=&
 \alpha_1\frac{\partial L  }{\partial A }   
 +  \alpha_2 \frac{\partial L }{\partial B }  
 +  \alpha_3 \frac{\partial L }{\partial \phi  } 
 +  \alpha_4 \frac{\partial  L}{\partial R } 
 \nonumber
 \\
 &&
 + \alpha_1'\frac{\partial L  }{\partial A' }   
 +  \alpha_2' \frac{\partial L }{\partial B' }  
 +  \alpha_3' \frac{\partial L }{\partial \phi ' } 
 +  \alpha_4' \frac{\partial  L}{\partial R' } 
\\
  &=& \alpha_1 \frac{B'^2 }{2 \kappa^2 B }
 \nonumber
 + \alpha_2 \left( \frac{-AB'^2 }{2 \kappa^2 B^2 }-\frac{R }{2 \kappa^2  }+ Y(R)\phi '^2\right)
 \nonumber
 \\
 &&
 +\alpha_4 \left( \frac{-B }{2 \kappa^2  }+BY_R(R)\phi '^2  \right)+\alpha_1'\frac{B'}{\kappa^2 }+\alpha_2'\left( \frac{AB'}{\kappa^2 B }+\frac{A'}{\kappa^2  }\right) 
 \nonumber
 \\
 &&
 +\alpha_3'2BY(R)\phi ' =0
 \end{eqnarray}
  where $\alpha_i=\alpha_i(A,  B , \phi  ,R)$. Here the derivatives $\alpha_i' $ can be written by the chain rule
 \begin{eqnarray}\label{556}
 \alpha_i '=\frac{\partial \alpha_i }{\partial A }A' +\frac{\partial \alpha_i }{\partial B }B'+\frac{\partial \alpha_i }{\partial \phi  }\phi '+\frac{\partial \alpha_i }{\partial R }R' 
 \end{eqnarray}
and the Lie derivative of the point-like Lagrangian gives us   the following  system of partial differential equations 
 \begin{eqnarray}\label{denk1}
 \frac{\alpha_1 }{2B }-\frac{\alpha_2 A }{2B^2 }+\frac{\partial \alpha_1 }{\partial B }+\frac{\partial \alpha_2 }{\partial B }\frac{A }{ B } &=&0,
\\
 \alpha_2 Y(R)+\alpha_4 B Y_R(R)+2 \frac{\partial \alpha_3 }{\partial \phi  } B Y(R) &=&0,\\
 \frac{\partial \alpha_1 }{\partial \phi  }\frac{1 }{ \kappa^2 }+\frac{\partial \alpha_2 }{\partial \phi  }\frac{1 }{ \kappa^2 }\frac{A }{ B }+2\frac{\partial \alpha_3 }{\partial B }B Y(R)&= & 0,
 \\
 \frac{\partial \alpha_1 }{\partial A }+\frac{\partial \alpha_2 }{\partial A }\frac{A }{ B }+\frac{\partial \alpha_2 }{\partial B}&=&0,
 \\
 \frac{\partial \alpha_2 }{\partial \phi  }\frac{1 }{ \kappa^2 }+2\frac{\partial \alpha_3 }{\partial A }B Y(R)&=&0,
 \\
 \frac{\partial \alpha_1 }{\partial R }+\frac{\partial \alpha_2 }{\partial R }\frac{A }{ B }&=&0,
 \\
 \frac{\partial \alpha_2 }{\partial A }=0 , \hskip 0.5 cm  \frac{\partial \alpha_2 }{\partial R }=0 , \hskip 0.5 cm  \frac{\partial \alpha_3 }{\partial R }B Y(R)&=&0,
\\
 \alpha_2 R+\alpha_4 B &=& 0.
 \end{eqnarray}
A solution to the system for an arbitrary $Y(R)$ function  can be found as 
\begin{eqnarray}
\alpha_1=\frac{c_1 }{\sqrt{B}  } , \hskip 0.5 cm  \alpha_2=0, \hskip 0.5 cm \alpha_3=c_2, \hskip 0.5 cm \alpha_4=0  \;.
\end{eqnarray}
Here $c_1, c_2 $ are  arbitrary constants. Then the $X$ vector field can be found as
\begin{eqnarray}
X = \frac{c_1}{\sqrt{B} } \frac{\partial}{ \partial A} - \frac{c_1}{2B^{3/2} } \frac{\partial}{ \partial A'}  + c_2 \frac{\partial }{\partial \phi } \ 
\end{eqnarray} 
and the  the  constant of motion  (\ref{canstantofmotion}) becomes 
\begin{eqnarray}
\Sigma_0 =\frac{2c_1}{\kappa^2}+2c_2 B Y(R)\phi ' \ . \label{Conserved charge} 
\end{eqnarray}

 \section{Euler-Lagrange equations} \label{model}
In order to determine the non-minimal function and the metric functions, 
we calculate the Euler-Lagrange equations from
\begin{eqnarray}
\frac{d}{dr} (\frac{\partial L}{\partial q'^i})-\frac{\partial L}{\partial q^i}=0
\end{eqnarray} 
 for the Lagrangian  (\ref{553}).  Then we obtain the following differential equations for $A,B,\phi ,R$, respectively:
\begin{eqnarray}\label{Bdif}
B''-\frac{B'^2}{2 B}=0   , \\
\frac{A'B'}{B}+\frac{AB''}{B}-\frac{AB'^2}{2B^2}+A''+\frac{R}{2}- Y(R)\phi '^2 \kappa^2=0,    \label{asildif}
\\
 (B\phi ' Y(R))'=0 \label{Maxwelldif},
\\
Y_R(R)\phi '^2=\frac{1}{2 \kappa^2  }  \ . \label{condition}
\end{eqnarray}
From  (\ref{Bdif})  and  (\ref{Maxwelldif}) we find 
\begin{eqnarray}
  B(r) = b_1(r+ b_2)^2, \hskip 1 cm \label{Br} 
  Y(R)\phi ' = \frac{q}{B},
 \end{eqnarray}
 where $q$ is an integration constant and  it corresponds to the electric charge of the source.     We note that the condition   (\ref{condition}) can be found by taking the derivative of equation (\ref{asildif}) with respect to $r$ as in \cite{Sert2016regular}.  Then we have only  the following differential equation  (\ref{asildif})  to solve
 \begin{eqnarray}
 \frac{A''}{2} - \frac{AB'^2}{4B^2} + \frac{1}{B} - \kappa^2 Y(R) \phi '^2 =0 \label{asildif2}
 \end{eqnarray}
 which is same with the differential equation  obtained from the field equations of the model in \cite{DereliSert20112,Sert2012Plus,Sert2013MPLA,Sert2016regular} for $B=r^2$. Thus we show that  these two different methods give the  same  differential equation (\ref{asildif2}). The conserved charge (\ref{Conserved charge})  of the model turns out to be
 \begin{eqnarray}
\Sigma_0 =\frac{2c_1}{\kappa^2}+ 2qc_2 \;
\end{eqnarray}

 for the   Noether symmetry. We see that the conserved quantity  involves the gravitational coupling constant $\kappa^2$ and the electric charge of the system $q$.   We also  calculate the energy function 
 from  \begin{eqnarray}
  E_L=q'^i(\frac{\partial L}{\partial q^i})-L
 \end{eqnarray}
 and find 
\begin{eqnarray}\label{energyfunction}
E_L=\frac{BR}{2 \kappa^2  }+ \frac{AB'^2}{2 \kappa^2 B} +\frac{A' B'}{ \kappa^2 } +BY(R)\phi '^2-\frac{2}{ \kappa^2 }\ .
\end{eqnarray}
By substituting the Ricci scalar (\ref{R}) in the energy function (\ref{energyfunction}), we find that the function is equal to zero,  since equation   (\ref{energyfunction}) is nothing more than equation  (\ref{asildif2}).
Furthermore, we can choose $B=r^2$ without loss  of generality then  (\ref{asildif2}) becomes
\begin{eqnarray}\label{asildif3}
\frac{A''}{2} +  \frac{1-A}{r^2} -  \kappa^2 Y(R)\phi '^2=0  \ .
\end{eqnarray}

\subsection{Some New Solutions }

In order to obtain solutions of the  differential equation  (\ref{asildif3}),   we can choose the non-minimal function $Y(R) $ that determines the strength of the coupling and find the metric function $A(r) $ as a first method. Alternatively,   we can choose possible geometries which are asymptotically flat  and involve correction terms to the known Reissner-Nordstrom solution as a second method. Then we can find the corresponding non-minimal function $Y(R)$.
Here we consider the second method
and we  take the following metric function  with the  Yukawa-like correction term 
\begin{eqnarray}\label{fonk1}
 A_1(r)=1-\frac{2M}{r  }  + \frac{q^2}{r^2  } - a\frac{e^{-r}}{ r^2   }
\end{eqnarray} 
which gives the solution 
\begin{eqnarray}\label{phi1}
\phi (r) &=& \frac{q}{r} - \frac{a e^{-r}(r+4)  }{qr}  \; ,\\
Y(r) &=&  \frac{8q^2/\kappa^2 }{4q^2   -  a (r+2)^2 e^{-r} } \label{Y1}
\; .
\end{eqnarray}
Then the Ricci scalar becomes  $R=\frac{a}{e^r r^2}$ for the metric function and by taking the inverse function $r=2W (x)$, we can re-express the non-minimal function (\ref{Y1}) in terms of $R$  as  
\begin{eqnarray}
Y(R)= \frac{8q^2/ \kappa^2} { 4q^2 - 4(W(x) + 1 )^2 e^{ -2W(x)} } 
\end{eqnarray}
where $W(x)$ is the Lambert function  with $x=\sqrt{ \frac{a}{4 R} }$. 	

Secondly, we  choose another metric function with the  Yukawa-like correction  term
\begin{eqnarray}\label{fonk2}
A_2(r)=1 - \frac{2M}{r  }  +\frac{q^2}{r^2  } -  (1 +\frac{4}{r} + \frac{6}{r^2} )ae^{-r} 
\end{eqnarray}
which is also asymptotically flat. Then we obtain the following solution   
\begin{eqnarray}\label{phi2}
\phi (r) & =& \frac{q}{r} - \frac{  a(r^3 +6r^2 +18r + 24  )e^{-r} }  {qr} \;, \\
Y(r) & =& \frac{8q^2/\kappa^2 }{ 4q^2 - a(r^4 +4r^3 +12r^2 +24r +24 )e^{-r}   } \ .
\end{eqnarray}
We calculate the Ricci  scalar  for the second metric as $R=ae^{-r}$ and the inverse function  $r=lnx $ with $x= \frac{a}{R}$. Then the non-minimal function becomes 
\begin{eqnarray}
Y(R)  =  \frac{8q^2/\kappa^2 }{ 4q^2 - (ln^4x +4ln^3x +12ln^2x +24lnx +24 )R   } \ \  .  
\end{eqnarray}
\subsection{Some Thermodynamic Properties of the Solutions}

The above metric functions (\ref{fonk1}) and (\ref{fonk2}) may describe a naked singularity without horizon or a black hole with one horizon or two horizons which are  called event horizon and Cauchy horizon depending on the choice of the parameters. In the cases with  event horizon $r=r_h$, the Hawking temperature is defined by
\begin{eqnarray}
T= \frac{A'(r_h)}{4\pi}
\end{eqnarray}
and the temperatures  can be  found 
\begin{eqnarray}
4\pi T_1 &=& \frac{a(r_h+2)}{r_h^3 e^{r_h}} +\frac{2M}{r_h^2} - \frac{2q^2}{r_h^3}\; ,
\\
4\pi T_2 &=& \frac{a(r_h^3 +4r_h^2 +10r_h +12) }{r_h^3e^{r_h} }) +\frac{2M}{r_h^2} - \frac{2q^2}{r_h^3} \; 
\end{eqnarray}
 for the above metric functions (\ref{fonk1}) and (\ref{fonk2}). We give the variation  of the temperatures  with  the event horizon radius $r_h$  in Fig. 1 for this non-minimal model and  the Reissner-Nordstrom case.
  \begin{figure}[t]{}
	\centering
	\subfloat[]{ \includegraphics[width=0.4\textwidth]{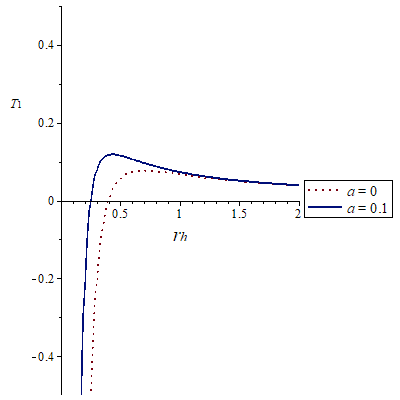} }
	\subfloat[  ]{ \includegraphics[width=0.4\textwidth]{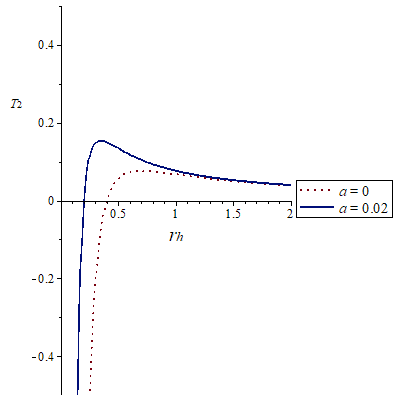}} 
	\\
	\parbox{6in}{\caption{{{\small{  The Hawking temperature versus the event horizon radius with $q=0.4$, the dotted curve is the Reissner-Nordstrom case and the solid curve is  the non-minimal  model.  
	}}}}}
\end{figure} 
By using the entropy of black hole   $S= \pi r_h^2$,
we calculate the heat capacity  from 
\begin{eqnarray}
C= T\left( \frac{\partial S }{\partial T}\right)_q
\end{eqnarray} for the above two metric functions and obtain 
\begin{eqnarray}
C_1& =& \frac{ 2 \pi r_h^2 \left(  ( r_h^2 -  q^2  )e^{r_h}   +     a ( r_h + 1  ) \right)   }{ ( 3q^2  - r_h^2 )e^{r_h} - a(r_h^2 +3r_h +3  ) } \; ,
\\
C_2&=& \frac{ 2 \pi r_h^2 \left(  ( r_h^2 -  q^2  )e^{r_h}   +     a ( r_h^3  +3r_h^2 +6r_h  + 6  ) \right)   }{ ( 3q^2  - r_h^2 )e^{r_h} - a(r_h^4  +3r_h^3  +9r_h^2 + 18r_h +18  ) } \; .
\end{eqnarray} 

Heat capacity gives us information about  thermal stability  intervals and phase transition points  of a black hole. Heat capacity must be  positive and finite  for a stable black hole.
 The points where heat capacity is zero give us  the  Type-1 instability and   the points  where heat capacity diverges  give us Type-2 instability points  which correspond to  the second order phase transition for  a black hole.    
We plot also the heat capacity   versus the horizon radius $r_h$  for the  Reissner-Nordstrom solution and the non-minimal model in Figure 2 with different ranges to  see  these points clearly. Numerically, we can find   upper bounds for the non-minimal parameter  $a$ as  $a=0.16$  for $A_1$ solution  and   $a=0.026$  for $A_2$ solution with $q=0.4$, to have a stable black hole.  Furthermore,
the type-2 instability points  decrease from $0.7$ to $0$, while $a$ increases from $0$ to the upper bounds, respectively. Moreover, these upper bounds of the parameter $a$  can  increase  to higher values as the electric charge increases.

  \begin{figure}[t]{}
	\centering
	\subfloat[ $  C_1$  for $0\leq r_h \leq 1.5$   ]{ \includegraphics[width=0.4\textwidth]{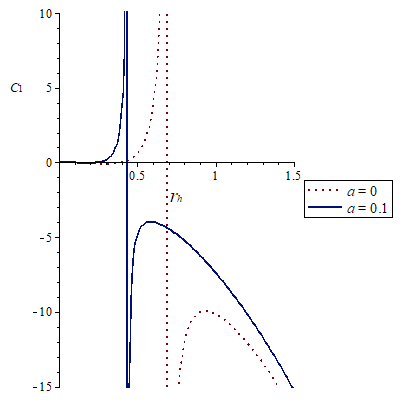} }
	\subfloat[  $C_2 $  for  $0\leq r_h \leq 1.5$   ]{ \includegraphics[width=0.4\textwidth]{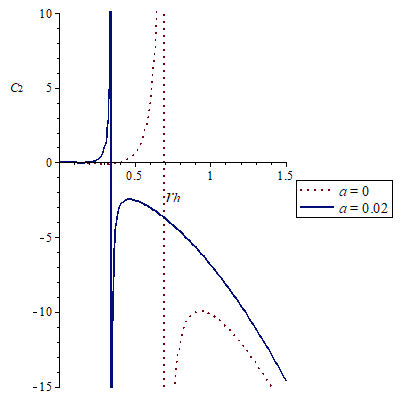}}\\ 
		\subfloat[ $  C_1$  for $0\leq r_h \leq 0.5$ ]{ \includegraphics[width=0.4\textwidth]{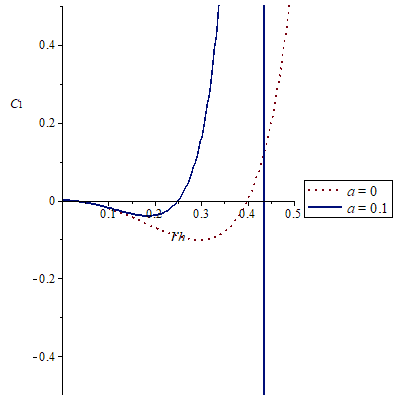} }
	\subfloat[  $  C_2$  for $0\leq r_h \leq 0.5$ ]{ \includegraphics[width=0.4\textwidth]{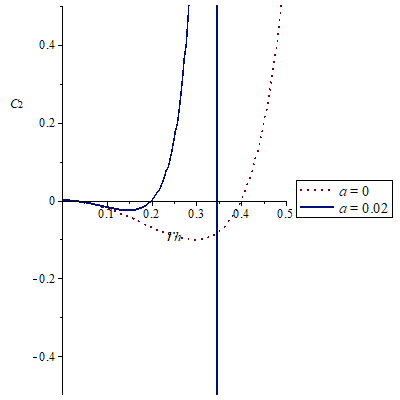}} 
	\\
	\parbox{6in}{\caption{{{\small{  The heat capacity versus the event horizon radius with $q=0.4$, the dotted curve corresponds to  the Reissner-Nordstrom case and the solid curve  to  the non-minimal  case.
	}}}}}
\end{figure}

On the other hand, the first law of thermodynamics is given by
\begin{eqnarray}
dM = TdS +\phi dq\end{eqnarray}for a non-rotating black hole with mass $M$  and electromagnetic potential $\phi$.  It is interesting to show that the mass $M$ in the first law   can be expressed by the Smarr formula \cite{Smarr1973}
\begin{eqnarray}
M= 2TS + \phi Q
\end{eqnarray} for the Reissner-Nordstrom black hole  in  the Einstein-Maxwell theory.

Furthermore, the Smarr relation can be also obtained from the Komar integral \cite{Komar1959,Breton2005,Balart2017,Mazhari2019}  with a correction term as
\begin{eqnarray}\label{Smarr}
M= 2TS + q\phi  -\frac{1}{2} \int \tau  dV  
\end{eqnarray} 
where $\tau $ is the trace of energy-momentum tensor obtaining from the gravitational field equation 
\begin{eqnarray}
G_a = \kappa^2 \tau_a\ 
\end{eqnarray}
 and it   can  be  related with the work density   \cite{Hayward1998}.
In the minimal Einstein-Maxwell case, this relation (\ref{Smarr}) is automatically satisfied, since 
the trace of  Maxwell energy-momentum tensor   is zero. 
In contrast to the Einstein-Maxwell theory, the non-minimally coupled $Y(R)F^2$ theory has a non-vanishing trace of energy-momentum tensor. By taking $\kappa^2 = 8\pi $,	  the trace  is found 
		
\begin{eqnarray}
\tau =\frac{1}{8\pi } *(G_a\wedge e^a) = \frac{R}{8\pi}= - \frac{1}{8\pi}[A'' +\frac{4A'}{r} +\frac{2}{r^2}(A-1) ]\ .
\end{eqnarray}
In order to investigate whether these  metric functions (\ref{fonk1}) and (\ref{fonk2}) satisfy the Smarr formula  we 
 calculate the correction term as
\begin{eqnarray}
\int_V \frac{\tau}{2} dV = \frac{1}{16\pi}\int_{r_h}^\infty R(r)4\pi r^2 dr=  \frac{M}{2} +\frac{r_h^2 A'(r_h)}{ 4} -\frac{r_h}{2}
\end{eqnarray}
for the  the metric  functions. Then we firstly consider the electric potential (\ref{phi1}) at the event horizon 
\begin{equation}
\phi = \int_{r_h}^\infty E dr =   \frac{ q}{r_h}- a\frac{e^{-r_h}(r_h+4)}{4qr_h} \ .
\end{equation}
Thus  the mass obtained  from the Smarr formula  (\ref{Smarr}) turns out to be
\begin{equation}
M=\frac{q^2+r_h^2 -ae^{-r_h}}{2r_h}
\end{equation}
and it is equal to the mass  obtained from  $A(r_h) = 0$.
   The Smarr formula (\ref{Smarr}) is also satisfied for the second solution (\ref{phi2}) similarly,   and    the  mass  is found 
\begin{equation}
M=\frac{q^2+r_h^2  -a(r_h^2 +4r_h + 6 ) e^{-r_h}}{2r_h} \ .
\end{equation}

\section{Conclusion}

In this study,  we have considered Noether symmetry approach to find    spherically symmetric, static   solutions of  the non-minimally coupled $Y(R)F^2$ theory. By considering the point-like Lagrangian of the  $Y(R)F^2$ theory with spherical symmetry, we  have found a vector field, satisfying the Noether symmetry condition,
and the corresponding conserved  quantity for any $Y(R)$ function.  We have  also derived Euler-Lagrange equations from the point-like Lagrangian. Then we have shown  that these equations are same with the equations derived from the field equations of the non-minimal model.

 We have  also given two exact asymptotically flat solutions  and the  corresponding non-minimal model. Then we have investigated   some thermodynamic properties of these solutions such as the Hawking temperature and the heat capacity to determine the thermal  stability intervals of  the solutions. 
 Furthermore we have shown that  the solutions satisfy the the modified Smarr formula for the models with non-zero energy-momentum tensor.



\end{document}